\documentclass[
    ,final            
  ]
  {aipproc}
\usepackage{multicol}

\layoutstyle{6x9}

\newcommand{\myFigure}[3]{%
    \begin{center}\begin{minipage}[t]{\columnwidth}%
    \refstepcounter{figure}\vspace{1ex}%
    \includegraphics[width=#1\columnwidth,keepaspectratio]{#2}\ \\%
    \bf Figure \arabic{figure}:\ \rm #3
    \vspace{1ex}%
    \end{minipage}\end{center}}

\newcommand{\apj}{ApJ}
\newcommand{\apjl}{ApJL}

\newcommand{\aap}{A\&A}

\newcommand{\xte}{{\it RXTE}}
\newcommand{\sax}{{\it BeppoSAX}}

\newcommand{\epcs}{{\rm ergs\,cm^{-2}\,s^{-1}}}
\newcommand{\epc}{{\rm ergs\,cm^{-2}}}

\newcommand{\srca}{XTE~J1814$-$338}
\newcommand{\srcb}{SAX~J1808.4$-$3658}
\newcommand{\srcc}{XTE~J0929$-$314}
\newcommand{\srcd}{XTE~J1751$-$305}
\newcommand{\srce}{XTE~J1807$-$294}
\newcommand{\srcf}{IGR~J00291+5934}
\newcommand{\srcg}{HETE~J1900.1$-$2455}

\newcommand{\gtsim}{\lower.5ex\hbox{$\; \buildrel > \over \sim \;$}}
\newcommand{\ltsim}{\lower.5ex\hbox{$\; \buildrel < \over \sim \;$}}

\newcommand{\nodata}{$\ldots$}

\begin{document}

\title{Accretion-powered Millisecond Pulsar Outbursts}


\author{Duncan K. Galloway}{
  address={School of Physics, University of Melbourne, VIC 3010 Australia}
}

\begin{abstract}
The population of accretion-powered millisecond pulsars (AMSPs) has grown
rapidly over the last four years, with the discovery of six new examples
to bring the total sample to seven.  While the first six 
discovered are transients active for a few weeks every two 
or more years, the most recently-discovered source \srcg, has been active for
more than 8 months. We summarise the transient behaviour of the population
to estimate long-term time-averaged fluxes, and equate these fluxes to the
expected mass transfer rate driven by gravitational radiation in order to
constrain the distances. We also estimate an upper limit of 6~kpc to the
distance of \srcf\ based on the non-detection of bursts from this source.
\end{abstract}

\maketitle

\section{Introduction}

Each of the known AMSPs have now been well-studied in
followup observations, and a review of their observational properties can
be found in \cite{wij03a}.
Six of the seven are transients with outburst intervals of 2 or more
years.
However, activity of the most recently-discovered example, \srcg\
\cite{vand05a},
has continued long beyond the usual active interval for the other six
sources. The latest observation, on 2006 February 2, indicates that the
source is still at approximately the same flux level as it was
throughout the second half of 2005. The source is also unusual in that the
pulsations are not detected consistently while the source is active
\cite{kaaret05a}.

Here we present a summary of the 
outburst history of the 
accretion-powered 
MSPs, 
in order to compare the long-term accretion rates.
We also introduce a new method which can give an upper limit on the
distance for sources where no bursts have been detected.

\section{Observations and Analysis}

We analysed observations of the 
AMSPs
made with the {\it Rossi X-ray Timing Explorer}\/ (\xte). 
We used measurements of the persistent flux and peak flux of thermonuclear
(type I) bursts (where available)
tabled in the catalog of Galloway et al. (2006a, in preparation).
The data were analysed with {\sc lheasoft} version 5.3, released 2003
November 17.
The persistent flux was measured by averaging the integrated flux from the
best-fit absorbed blackbody plus power-law model in the energy range
2.5--25~keV, to spectra extracted separately for each PCU.
We used a bolometric correction to the 2.5--25~keV flux based on
absorbed comptt model fits to combined PCA and HEXTE
data.
The X-ray colors for individual AMSPs were relatively constant over each
outburst, and so we
adopted a constant correction for each source:
\srca, 1.86;	
\srcb, 2.12;	
\srcc, 1.80;	
\srcd, 1.66;	
\srce, 1.57; and	
\srcf, 2.54.	

\section{Results}
\label{results}

We estimated the fluence for each outburst 
using public \xte\/ PCA and ASM measurements. 
For intervals where the outburst was not covered by PCA observations,
we integrated the ASM intensities 
instead, using a linear cross-calibration
between the PCA and the nearest 1-day average 2--10~keV ASM intensities.
\srce\ and \srcd\ lie towards the Galactic center, and a cataclysmic
variable is within the $1^\circ$ \xte\/ field-of-view centred on \srcf.
Thus, for those sources we subtracted out a baseline level of 10, 5 and
$5\times10^{-11}\ \epcs$ respectively, which we attribute to
contributions from diffuse background and/or unrelated field sources.
The fluences derived by this method, scaled to give the estimated
bolometric values, are listed in Table \ref{outbursts}.
We propagated the errors from the uncertainties on individual ASM/PCA
measurements.
We note that the calculated fluences were generally consistent with prior
estimates, to within the uncertainties.

The time-averaged accretion rate driven by angular momentum loss arising
from gravitational radiation from the binary is given by \cite[]{bc01}
\begin{equation}
  \dot{M}_{\rm GR} \gtsim 
      3.8\times10^{-11} \left(\frac{M_C}{0.1M_\odot}\right)^2
             \left(\frac{M_{\rm NS}}{1.4M_\odot}\right)^{2/3}
             \left(\frac{P_{\rm orb}}{2\ {\rm hr}}\right)^{-8/3}\
             M_\odot\,{\rm yr^{-1}}
 \label{mgr}
\end{equation}
where $M_C$ is the minimum companion mass, $M_{\rm NS}$ the neutron star
mass, and $P_{\rm orb}$ the binary orbital period.
Because pulse timing allows measurement only of the projected
semimajor axis $a_X\sin i$,
only a lower limit on $M_C$ is available.
Thus, on equating the time-averaged X-ray flux $\left<F_X\right>$ and
$\dot{M}_{\rm GR}$, we derived lower limits on the distance $d$ for the
interval prior to each outburst (Table \ref{outbursts}).

\begin{table}
\begin{tabular}{lcccccccc}
\hline
 \tablehead{1}{l}{b}{Source} &
  \tablehead{1}{c}{b}{Outburst} &
  \tablehead{1}{c}{b}{Start\\(MJD)} &
  \tablehead{1}{c}{b}{Interval\\(yr)\tablenote{The epoch for the outburst
prior to the first known is assumed to be earlier than the first ASM
measurements (typically 1996 January 6 or MJD~50088).}} &
  \tablehead{1}{c}{b}{Fluence\tablenote{Bolometric fluence, in units of 
$10^{-3}\ \epc$}} &
  \tablehead{1}{c}{b}{$\left<F_X\right>$\tablenote{Estimated
time-averaged bolometric flux in units of $10^{-11}\ \epcs$.}} &
  \tablehead{1}{c}{b}{Distance\\limit (kpc)\tablenote{The values or
limits in parentheses are based on an assumed fluence for the outburst, in
the cases where the fluence of only one outburst has been measured with
any precision.}} 
\\
\hline
XTE J1807$-$294    & Feb 2003 & 52681 & $>7.1$ & $3.1\pm0.2$ & $<1.4$
  & 4.7 \\ 
XTE J1751$-$305    & Jun 1998 & 50978 & $>2.4$ & \nodata & ($<3.0$)
  & (6.2) \\ 
                   & Apr 2002 & 52363 & 3.8 & $2.3\pm0.3$ & 1.9 
  & (7.8) \\ 
XTE J0929$-$314    & Apr 2002 & 52376 & $>6.3$ & $5.4\pm0.3$ & $<2.7$ 
  & (3.6) \\ 
SAX J1808.4$-$3658 & Sep 1996 & 50333 & $>0.67$ & $7.7\pm0.6$ & $<36$
  & 1.4 \\ 
                   & Apr 1998 & 50911 & 1.58 & $5.2\pm0.5$ & 10 
  & 2.5 \\ 
                   & Jan 2000 & $\approx51547$ & 1.74 & $5.4\pm1.7$ & 9.8 
  & 2.6 \\ 
                   & Oct 2002 & 52559 & 2.8 & $6.2\pm0.4$ & 7.0 
  & 3.1 \\ 
                   & June 2005 & 53522 & 2.6 & $4.9\pm0.6$\tablenote{The
fluence for the June 2005 outburst of \srcb\ was estimated from the ASM
observations alone, since no public PCA data were available}
  & 5.9 & 3.4 \\ 
IGR J00291+5934    & Nov 1998 & 51143 & $>2.9$ & \nodata & ($<1.8$)
  & (4.3) \\ 
                   & Sep 2001 & 52163 & 2.8 & \nodata & (1.8) 
  & (4.2) \\ 
                   & Dec 2004 & 53341 & 3.2 & $1.63\pm0.16$ & 1.6 
  & 4.5 \\ 
XTE J1814$-$338    & Jun 2003 & 52789 & $>7.4$ & $2.99\pm0.12$ & $<1.3$ 
  & 10.5 \\ 
\hline
\caption{Outburst properties and distance limits for the
millisecond X-ray pulsars
 \label{outbursts} }
\vspace{-12pt}
\end{tabular}
\end{table}

For the three sources with thermonuclear bursts, independent estimates of
the distance can be made from the peak burst flux. Based on bursts
observed by {\it BeppoSAX}, \cite{zand01} estimated $d=2.5$~kpc for \srcb\
or up to 3.3~kpc for a pure He burst. Similarly, \cite{kawai05}
estimated $d=5$~kpc for \srcg\/ based on a burst observed with
{\it HETE-II}. While the brightest burst from \srca\ was not conclusively
shown to exhibit radius-expansion, the implied 
$d<8$(10)~kpc for $X=0.7$(0.0) \cite{stroh03a}. Based on these distances, or
the limits from Table \ref{outbursts} for those sources with no detected
bursts, we estimated the long-term averaged $\dot{M}$ for each of the
AMSPs (Fig. \ref{mdot}).

We note that the measured outburst fluences and intervals for
\srcb\ indicate that $\left<F_X\right>$ (and hence $\dot{M}$)
is decreasing steadily.
Since the distance
depends on the flux only to the $-1/2$ power,
the derived limit varied only by 30\%, up to a
maximum of 2.9~kpc (Table \ref{outbursts}).
\srcf\ is the only other source for which multiple outbursts have
been identified, and the increasing outburst interval suggests
that 
$\dot{M}$ may also have been decreasing with time.

Markedly different behaviour has been exhibited by \srcg\ since its
discovery in 2005 June \cite{vand05a}.  Although the source was too close to the sun for
observations during 2005 December and 2006 January, activity has
apparently continued for more than 8 months.
While the estimated $\dot{M}$ in outburst
(based on the approximately constant flux level of
$\approx9\times10^{-10}\ \epcs$ since 2005 June 14 \cite{gal05d})
is just 2\%~$\dot{M}_{\rm Edd}$ (for $d=5$~kpc),
continuing activity would make this the AMSP with the highest average
$\dot{M}$ by far (Fig. \ref{mdot}).

 \paragraph{Distance upper limits for non-bursting AMSPs}
 \label{nobursts}

While thermonuclear bursts have not been detected from 
four of the AMSPs, we expect that this is because they have 
been missed in data gaps
rather than being absent altogether, as in (e.g.) the high-field pulsars. 
\xte\/
is in a low-Earth orbit with a period of
$\approx90$~min, and suffers regular interruptions when observing
most of the sky due to Earth occultations, 
as well as observations of other
sources and passages through regions of high particle density, which
introduce additional gaps.
The AMSPs be arbitrarily
distant, because the implied $\dot{M}$ would exceed the Eddington
limit; we may however infer a lower limit, at which point the 
implied $\dot{M}$ would be high enough to produce sufficiently frequent
X-ray bursts that it would be highly improbable that they would all be
missed by the \xte\/ observations.

The key factors to determine the likelihood of burst
detection are the time density of observations (duty
cycle) and the underlying burst rate,
which depends in turn on $\dot{M}$ and the H-fraction in the accreted
fuel, $X_0$.
Three of the four AMSPs
in which no bursts have been detected are in ``ultracompact'' binaries
with $P_{\rm orb}\approx43$~min. The Roche lobes in 
such tiny binaries
cannot contain a
main-sequence companion, indicating that the mass 
donors are evolved and
(probably) H-poor (e.g. \cite{prp02}). 
The expected burst recurrence 
times 
are thus
very long
due to the absence of heating from persistent 
H-burning between 
\setlength{\columnsep}{10pt}
\begin{multicols}{2}
  \myFigure{0.95}{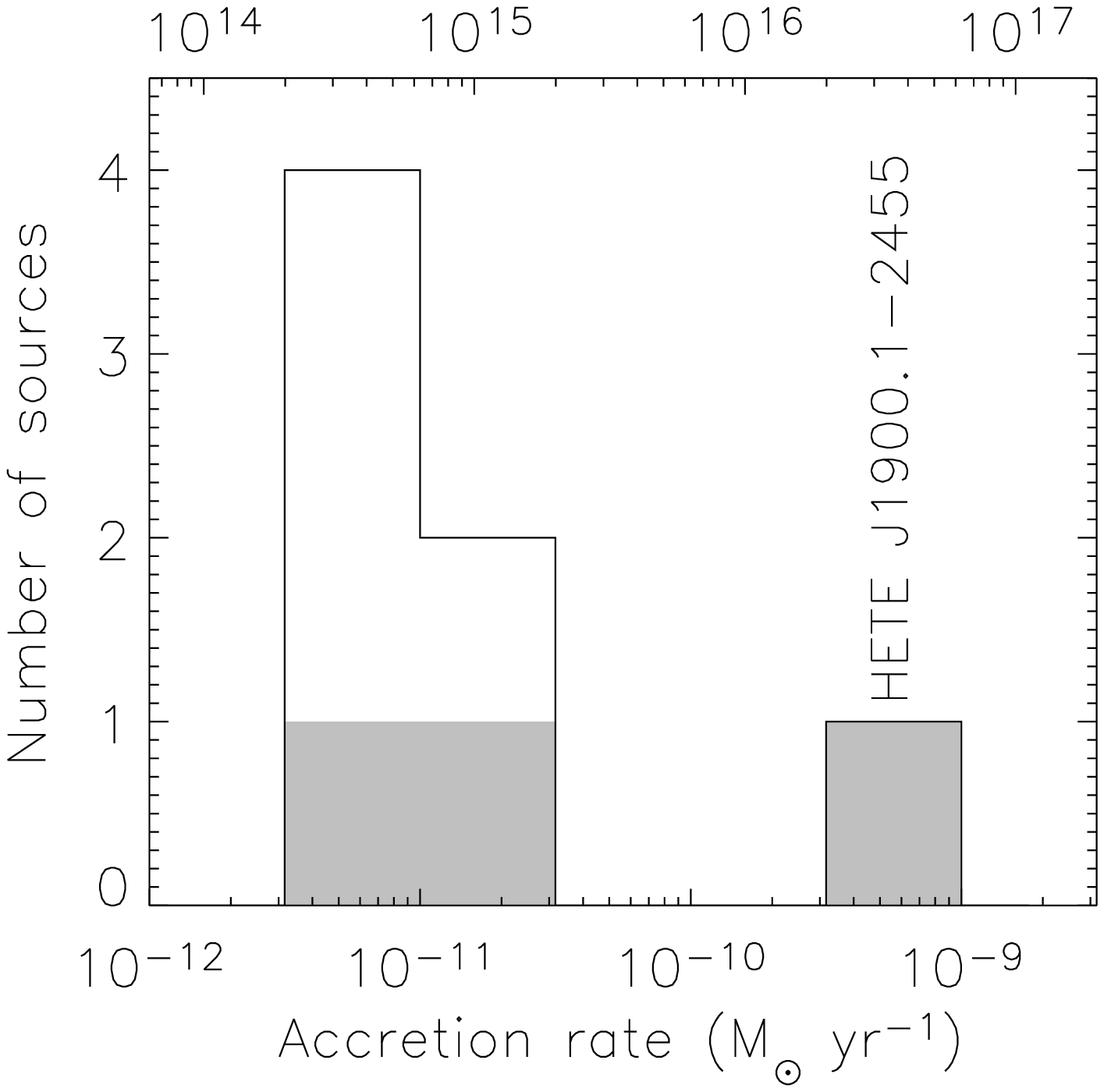}{Distribution of time-averaged $\dot{M}$ for the
AMSPs. The shaded histogram shows the distribution for sources with
distances measured from the peak flux of thermonuclear bursts; the other
values are from lower limits on $d$. The estimated $\dot{M}$ for \srcg\ in
outburst, based on the reported source flux 
and distance,
is indicated. The top $y$-axis is in units of g~s$^{-1}$.
 \label{mdot} }
  \myFigure{0.95}{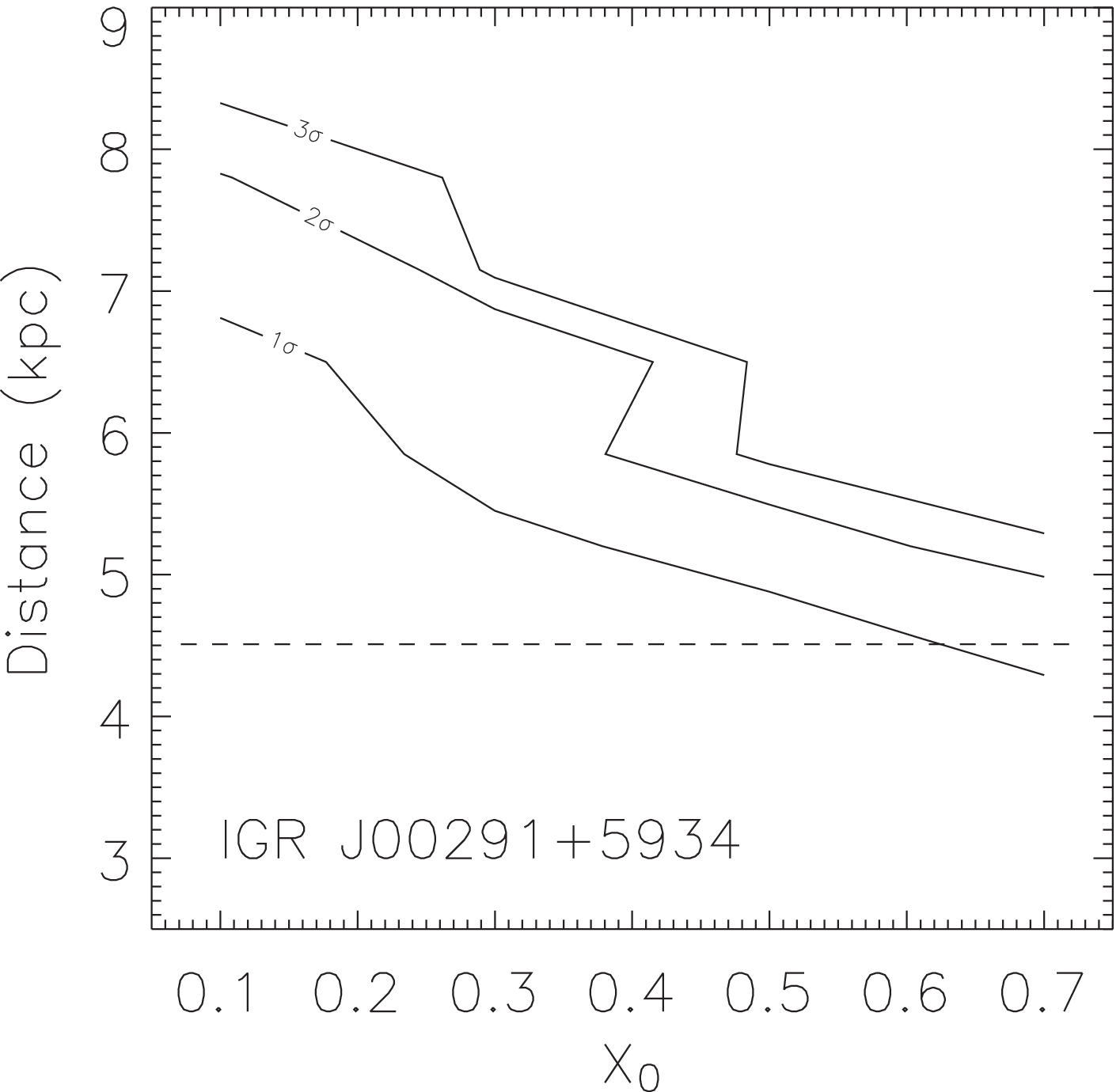}{
Combined distance-hydrogen fraction limits based on the
December 2004 outburst of \srcf. The dashed line indicates the lower limit
on the distance based on measurements of the time-averaged flux (Table
\ref{outbursts}. The contours indicate the estimated likelihood that all
the bursts were missed for each combination of $X_0$ and $d$.
 \label{ratelim} }
\end{multicols}
\vspace{-18pt}
\noindent
the bursts
(e.g. \cite{cb00}).
The combination of very low expected burst rates for these sources, and
low duty cycles for the PCA observations (e.g. 6.6\% for \srce) makes
it difficult to constrain the distances.
The duty cycle for the PCA observations of \srcf, on the other hand,
was higher, at 27\%. Since the mass donor in this source is also thought
to be H-rich, we expect the highest burst rate of all four sources with no
detected bursts, 
and thus is the least likely to have missed
all the bursts.

We generated plausible burst sequences for \srcf\ based on
the cubic-spline interpolated flux evolution measured during the 2004 December
outburst,
with burst ignition conditions calculated as in
\cite{cb00}, to which we refer the reader for further details.
We adopted a grid of distances beginning at the lower limits in Table
\ref{outbursts}.
We assumed a $1.4\ M_\odot$ neutron star with radius
$R=10\ {\rm km}$, giving a surface gravity $g=(GM/R^2)(1+z)=2.45\times
10^{14}\ {\rm cm\ s^{-2}}$ and redshift $1+z=1.31$.
We generated
$10^4$ burst sequences for each value of $d$ and 
$X_0=0.1$, 0.3, 0.5 and 0.7. We fixed $Z_{\rm
CNO}=0.016$ (equivalent to solar metallicity) throughout.
We varied the start time of the first burst evenly within the
first predicted burst interval, and also 
introduced a modest degree of scatter on the burst times,
with a standard deviation of 0.13~hr
We then
checked how many of the predicted burst times fell within the intervals during
which the PCA was observing the source.
We interpreted
the fraction of trials which resulted in one or more detected bursts, as
the probability that we could reject that set of parameters.

At the lowest value of $X_0=0.1$, the predicted burst rate was
sufficiently low that the likelihood of
missing any bursts present was high, even for distances as
large as 8~kpc.
However, for higher values of $X_0$, the higher $\dot{M}$ implied by
such large distances made it increasingly unlikely that we would have
missed all the bursts. Thus, the likely distance limit became smaller.
Since we expect the mass donor in \srcf\ is H-rich, we expect a source
distance of no more than 6~kpc (Fig. \ref{ratelim}).

\section{Discussion}
\label{disc}

We have derived distances or limits to the known AMSPs via 
analysis of \xte\/ observations.
Based on the outburst history of the AMSPs, the next outburst expected is
from \srcd, early in 2006.
In both cases where more than two
outbursts are known (\srcb\ and \srcf), we find evidence that the
long-term averaged flux is decreasing.
For the two transients with independent constraints on the distance from 
the peak flux of photospheric radius-expansion bursts, \srcb\ and \srca,
the
maximum lower
distance limit derived from equating the average flux and 
$\dot{M}_{\rm GR}$
is 
just above
the distance range derived from the bursts.
We also derived an upper limit on the distance to \srcf\ of 6~kpc,
based on
the predicted burst rate and the duty cycle for the \xte\/ observations.
Given a sufficiently high
observational duty cycle, this method may be used to derive distance
limits on other LMXBs where bursts have not been detected.

Finally, we note the discovery of \srcg, the first ``quasi-persistent''
AMSP. 
As of 2006 February this source is still active, more than 8
months after its discovery. Although the estimated 
$\dot{M}\approx2$\%~$\dot{M}_{\rm Edd}$ is significantly lower than the peak
reached by most of the other transient AMSPs, the fact that activity is
continuous indicates that the long-term time-averaged $\dot{M}$ of
this source may exceed all the others. In that case, \srcg\ is the best
candidate for the detection of gravitational waves from an AMSP.

\begin{theacknowledgments}
We thank Lars Bildsten and Philip Podsiadlowski for useful discussions.
This research has made use of data obtained through the 
HEASARC Online Service, provided by
NASA/GSFC. This work was supported in part by
the NASA 
LTSA program under grant NAG 5-9184.
\end{theacknowledgments}

\IfFileExists{\jobname.bbl}{}
 {\typeout{}
  \typeout{******************************************}
  \typeout{** Please run "bibtex \jobname" to optain}
  \typeout{** the bibliography and then re-run LaTeX}
  \typeout{** twice to fix the references!}
  \typeout{******************************************}
  \typeout{}
 }

\end{document}